\documentclass[aps,prl,twocolumn,showpacs,preprintnumbers,amsmath,amssymb,superscriptaddress,floatfix]{revtex4}%
\usepackage{epsfig}%
\usepackage{dcolumn}
\usepackage{bm}
\begin{document}
\title{\Large\bf\boldmath 
Intra-unit-cell Magnetic Order in Stoichiometric La$_2$CuO$_4$ 
}%
\author{Vyacheslav G.~Storchak}
\email{mussr@triumf.ca}
\affiliation{National Research Centre ``Kurchatov Institute'',
 Kurchatov Sq.~1, Moscow 123182, Russia}
\author{Jess H.~Brewer}
\affiliation{Department of Physics and Astronomy,
 University of British Columbia,
 Vancouver, BC, Canada V6T 1Z1}
\author{Dmitry G.~Eshchenko}
\affiliation{Bruker BioSpin AG, Industriestrasse 26, 8117 F{\"a}llanden, Switzerland}
\author{Patrick W.~Mengyan}
\affiliation{Department of Physics, Texas Tech University, 
 Lubbock, Texas 79409-1051, USA}
\author{Oleg E.~Parfenov}
\affiliation{National Research Centre ``Kurchatov Institute'',
 Kurchatov Sq.~1, Moscow 123182, Russia}
\author{Andrey M.~Tokmachev}
\affiliation{National Research Centre ``Kurchatov Institute'',
 Kurchatov Sq.~1, Moscow 123182, Russia}
\author{Pinder Dosanjh}
\affiliation{Department of Physics and Astronomy,
 University of British Columbia,
 Vancouver, BC, Canada V6T 1Z1}
\author{Sergey N. Barilo}
\affiliation{Institute of Solid State and Semiconductor Physics, 
Minsk 220072, Belarus}

\vfil 
\date{1 May 2014} 
\begin{abstract}
Muon spin rotation measurements supported by magnetization experiments  
have been carried out in a stoichiometric high-$T_c$ parent compound 
La$_2$CuO$_4$ in 
a temperature range from 2~K to 340~K 
and 
in transverse magnetic fields up to 5~T. 
Along with the antiferromagnetic local field, 
muon spin rotation spectra 
indicate presence of an additional source of magnetic field on the muon.
The characteristic splitting of about 45~G coming from this additional magnetic field
is consistent with spontaneous circulating currents model of Varma.
\end{abstract}
\vfil
\maketitle%
The idea of a pseudogap (PG) has recently 
played a key role in our understanding of 
strongly correlated electron systems --- 
particularly those exhibiting high-$T_c$ superconductivity (SC).
Although some basic phenomenology 
of high-$T_c$ copper oxide superconductors  --- 
electron pairs with nonzero angular momentum,
an 
exchange mechanism arising from 
strong Coulomb interaction between the valence electrons,
and 
a remarkable departure from the 
standard 
Fermi liquid behavior
in the normal state --- is a matter of growing consensus,
the microscopic mechanism still remains unclear.
Understanding 
 the origin of the PG and 
its relation to high-$T_c$ SC
is considered 
an important step towards revealing this mechanism 
\cite{Norman2003,Tallon2001,Timusk1999}.

All of 
these cuprates, being hole (p) doped, have the same T-p phase diagram:
at zero doping they are antiferromagnetic (AFM) Mott insulators
while doping easily 
destroys the AFM and makes the system metallic.
In the overdoped regime, the normal state exhibits properties 
of a correlated 
Fermi liquid. 
By contrast, in the underdoped regime of the phase diagram, 
cuprates show features of a correlated metal 
exhibiting non-Fermi-liquid 
behavior. 
Their transport, magnetic and thermodynamic properties 
point towards 
strong reduction of the electronic density of states (DOS) 
below a temperature $T^*$, 
although the DOS does not reach zero at the lowest temperature 
and the system remains metallic \cite{Norman2003} 
--- hence the PG terminology.  
Angle-resolved photoemission spectroscopy (ARPES) 
\cite{Kaminski2002} 
suggests opening of a 
real gap 
in 
the 
one-particle excitation spectrum, 
supported by other 
spectroscopic techniques \cite{Timusk1999}.

As doping dependence of the SC gap follows $T^*$ rather than $T_c$, 
the 
PG has been considered as a precursor 
to the SC gap;
the PG phase being  
a {\sl disordered\/} state with broken phase coherence 
among preformed pairs, 
which condense below $T_c$ as soon as 
phase coherence is established \cite{Emery1995}.
A 
different approach is presented
by certain 
models which consider 
the PG state as an {\sl ordered\/} phase with a well defined 
order parameter and a related broken symmetry 
\cite{Varma1997,Varma2006,Chakravarty2001}. 
In this scenario, 
$T^*$ 
is the transition temperature to an ordered PG phase 
of orbital magnetic moments 
caused by spontaneous circulating currents (CC).  
The fluctuations associated with the broken symmetry 
are considered to be responsible for both the superconductivity, 
playing role of a pairing glue, and the 
non-Fermi-liquid behavior below 
$T^*$. 

To date, ARPES data 
are rather controversial:
an apparent spontaneous dichroism of ARPES spectra 
of Bi$_2$Sr$_2$CaCu$_2$O$_{8+\delta}$ \cite{Kaminski2002}
indicates a time-reversal symmetry breaking (TRSB) 
magnetic field in the PG phase --- a fingerprint of an ordered state.
Such dichroism, however, was not 
found in a later experiment 
\cite{Borisenko2004}.
The key experimental evidence for a novel ordered state 
comes from spin-polarized neutron scattering
experiments, which report 
commensurate magnetic peaks 
below $T^*$ in YBCO and Hg1201 
\cite{Fauque2006,Bourges2008,Li2008}, 
pointing to 
TRSB that, nevertheless, preserves lattice translation invariance,
as the nuclear and magnetic peaks in reciprocal space
are superimposed on Bragg reflections.
Sizeable magnetic moments (on the order of 0.1-0.2 $\mu_{\rm\scriptsize B}$) 
are reported in the PG state.

Positive muons, as local microscopic magnetic probes,
are especially sensitive to
any kind of magnetic order, 
manifested as 
a coherent muon spin oscillation 
with frequency proportional 
to the local magnetic field at the muon 
\cite{Brewer1994}.
However, muon spin rotation ($\mu $SR) experiments, 
which have convincingly demonstrated their 
sensitivity to TRSB fields in a number 
of weak magnetic systems, produced no evidence 
of magnetic order in the PG state. 
At first, spontaneous static magnetic fields 
were reported in YBCO \cite{Sonier2001}, 
which were later reinterpreted as being due to 
spatial charge inhomogeneities \cite{Sonier2002}. 
No TRSB is reported in La$_{2-x}$Sr$_x$CuO$_4$ 
(for $x$=0.13 and $x$=0.19, both within the PG regime)
where $\mu $SR experiments 
set an upper limit of $\sim 0.2$~G for any magnetic 
field at the muon site, while the expected TRSB local field 
is estimated to amount about 40~G \cite{MacDougall2008}.
This discrepancy between neutron and muon experiments 
has been attributed to screening of the 
charge density in the metallic-plane unit cells 
in the vicinty of the muon \cite{Shekhter2008}. 

On the other hand, orbital currents and 
associated magnetic moments may well 
be present in the limiting case 
of the underdoped regime --- 
insulating stoichiometric La$_2$CuO$_4$ ---  
should CC be an intimate feature of chemical bonding 
in the CuO$_2$ plane.
Recent 
studies of the phase diagram for 
high-$T_c$ cuprate SC suggest that CC loops 
do persist down to zero doping 
\cite{Weber2014}.

In this Letter, we present 
$\mu $SR spectroscopy of stoichiometric La$_2$CuO$_4$, 
in which the magnetic field at the muon should not be 
affected by charge density screening.
Our data indicate 
{\sl an additional source of magnetic field\/} 
at the muon site 
(over and above 
the previously known AFM field) 
consistent with Varma's model of circulating 
currents \cite{Varma1997,Varma2006} and polarized neutron experiments.

Single crystals of La$_2$CuO$_{4+x}$ grown 
from CuO flux 
were 
used for these 
studies. 
The crystal orientation, lattice parameters and mosaicity
(less than 0.05$^\circ$ along the $\hat{c}$-axis)
were 
determined by X-ray 
diffractometry. 
Surplus oxygen 
was 
removed by annealing in vacuum. 
The lattice parameters 
correspond 
to the low-temperature orthorhombic stoichiometric 
La$_2$CuO$_4$ ($Bmab$ space group) \cite{Parfenov2002}.
The N\'eel temperature $T_N=320$~K was measured by 
SQUID.  

Time-differential $\mu $SR experiments 
using 100\% spin-polarized positive muons 
were carried out on the M15 surface muon channel at TRIUMF
using the {\it HiTime\/} spectrometer.  
At low temperature, the zero-field (ZF) $\mu $SR spectra 
consist of two components (small and large amplitude), 
well known from 
previous studies and indicative of two 
inequivalent muon sites in AFM
La$_2$CuO$_4$. 
The N\'eel temperature and 
magnetic fields at the muon sites $B_\mu=428.7$~G 
(high-frequency, large amplitude component) 
and $B_\mu=111.8$~G (low-frequency, small amplitude component)
\cite{Storchak2014} 
are consistent with 
earlier studies. 

\begin{figure}[tb]
\begin{center}
\vspace*{1mm}
\includegraphics[width=\columnwidth,angle=0]{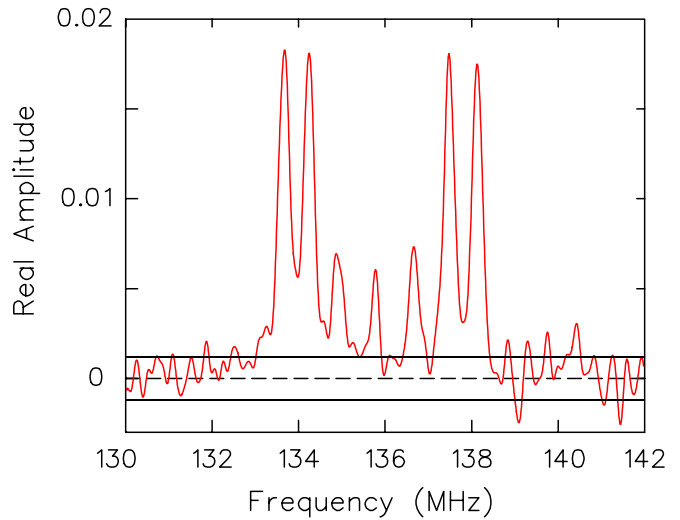}
\vspace*{-4mm}
\caption{\label{fig:La2CuO4-100K-1T-spec}
 Frequency spectrum of muon spin precession 
 in AFM stoichiometric La$_2$CuO$_4$ 
 in a transverse magnetic field of $H=1$~T at $T=100$~K.  
 The central line frequency coincides with that 
 detected in a CaCO$_3$ reference 
 sample. 
}
\end{center}
\vspace*{-8mm}
\end{figure}

\begin{figure}[b]
\begin{center}
\vspace*{-4mm}
\includegraphics[width=\columnwidth,angle=0]{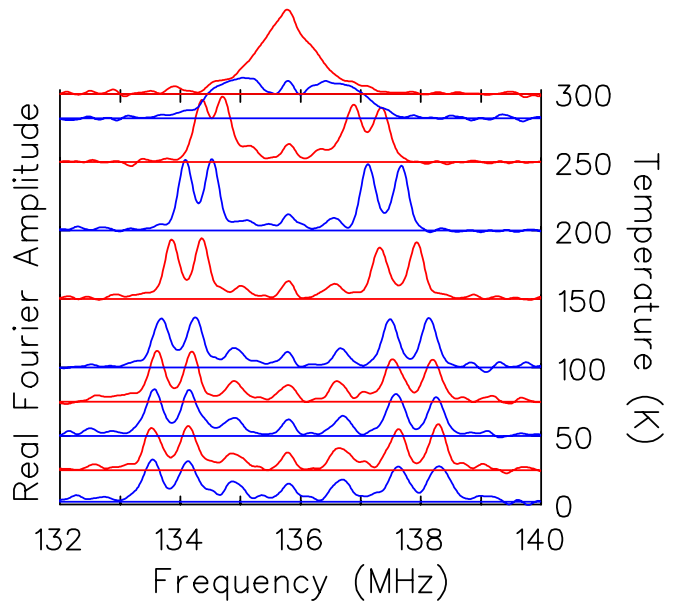}
\vspace*{-5mm}
\caption{\label{fig:La2CuO4-FFT1T-Tdep}
 Fourier transforms of the muon spin precession signal
 in AFM stoichiometric La$_2$CuO$_4$ 
 in an external magnetic field of $H=1$~T at different temperatures. 
 Again, the central line 
 corresponds to the background signal.  
}
\end{center}
\vspace*{-7mm}
\end{figure}

In high magnetic field $H$ applied 
transverse to the 
muon spin polarization and parallel to the 
$\hat{c}$-axis of the La$_2$CuO$_4$ crystal, 
$\mu$SR spectra exhibit 
at least 7 signal components (Fig.~\ref{fig:La2CuO4-100K-1T-spec}). 
The central line at about 135.6~MHz coincides with 
the single line detected in CaCO$_3$, 
a non-magnetic reference sample, 
and originates from
muons 
that 
miss the sample and 
stop in a non-magnetic environment. 
Two small peaks positioned 
around the central line 
correspond to the AFM splitting 
of the low-frequency (low-amplitude) signal 
observed in zero magnetic field. 
Our main interest is focused on the large-amplitude signals 
positioned 
symmetrically 
around the central line. 
We 
associate these signals with the high-frequency 
(large-amplitude) signal in zero magnetic field. 
The evolution of $\mu$SR spectra with temperature is presented 
in Fig.~\ref{fig:La2CuO4-FFT1T-Tdep}.

Within the AFM phase, one should expect 4 distinct signals 
in the high-field $\mu $SR spectra from two independent 
(and distinct) 
muon 
sites 
in the AFM host.  
Instead, we detect at least 6 peaks coming 
from muons stopped in the La$_2$CuO$_4$ crystal.
This may indicate that, apart from the AFM fields, there is 
{\sl an additional source of magnetic field on the muon\/} 
which causes the characteristic splitting of the 
large-amplitude signals (see 
Figs.~\ref{fig:La2CuO4-100K-1T-spec} 
and \ref{fig:La2CuO4-FFT1T-Tdep}). 
To determine the origin of this field, 
one needs 
a detailed interpretation of the 
$\mu $SR spectra based on the muon sites 
in La$_2$CuO$_4$. 

Identification of the muon stopping sites
from
ZF-$\mu $SR spectra 
has been 
discussed by many authors 
\cite{Hitti1990,
Sulaiman1994,Suter2003,Huang2012,Adiperdana2013}. 
However, information obtained 
from ZF spectra 
alone 
is 
insufficient 
and additional guidance is typically 
sought from first-principles 
electronic structure calculations. 
It is common 
to predict the muon stopping site to be 
in the vicinity of the apical oxygen 
atom, but the 
muon 
positions 
that are actually determined 
are quite different in a 
number of studies, 
which leads 
to much confusion. 
In contrast, 
high-field $\mu $SR measurements, which 
determine the magnitudes of the {\sl projections\/} 
(onto the directions of the applied field) 
of the local magnetic field at the muon, 
provide more precise information on the muon positions 
in La$_2$CuO$_4$  
(in combination with ZF data and the known crystal 
and magnetic structure of the system). 

At high temperature, the La$_2$CuO$_4$ 
crystal is 
tetragonal ($I4/mmm$ space group) 
\cite{Johnston1991,Kastner1998} 
with a unit cell formed by 2 formula units. 
A transition to the orthorhombic phase 
($Bmab$ space group) occurs at 530~K 
which doubles the unit cell. 
This structural transition is 
due to rotation of the central CuO$_6$ 
octahedra around the tetragonal axes (110) 
and antiphase rotations of 
octahedra in the neighbouring unit cells. 
At low temperature, the tilting 
of CuO$_6$ octahedra reaches 5$^\circ $ \cite{Johnston1991}. 
Stoichiometric La$_2$CuO$_4$ is a 
collinear AFM 
with four sublattices. 
The magnetic moment of a Cu atom 
amounts to 0.66$\pm $0.13~$\mu_{\rm\scriptsize B}$ 
and is directed along the diagonal 
of the CuO$_2$ plaquette \cite{Johnston1991,Scharpf1990}. 
A peculiar feature of magnetic ordering in 
La$_2$CuO$_4$ is the presence of a 
weak ($\sim 0.002$ $\mu_{\rm\scriptsize B}$ per Cu atom) 
ferromagnetic coupling of spins within 
CuO$_2$ layers \cite{Johnston1991,Thio1988} 
possibly 
originating from Dzyaloshinskii-Moriya 
exchange interactions and leading to 
a small tilting ($\sim 0.17^\circ$) 
of Cu magnetic moments \cite{Kastner1998,Thio1988}. 
These small moments are orthogonal to 
CuO$_2$ layers and have opposite directions 
in the neighbouring planes. 
Thus La$_2$CuO$_4$ is an antiferromagnet 
with hidden, weak ferromagnetism.

\begin{figure}[tb]
\begin{center}
\vspace*{1mm}
\includegraphics[width=\columnwidth,angle=0]{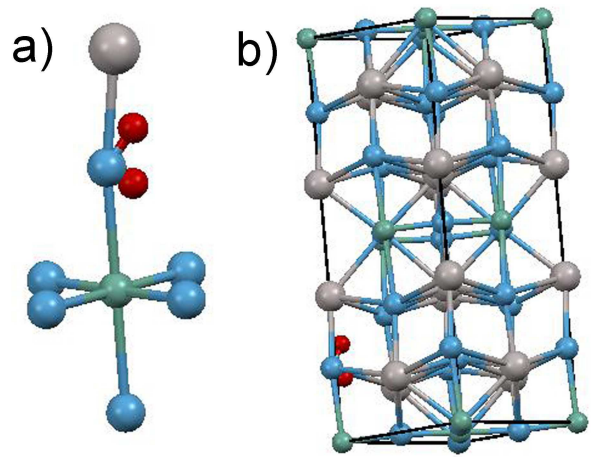}
\vspace*{-4mm}
\caption{\label{fig:La2CuO4-muons}
 Muon stopping sites as determined from dipole-field
 calculations and shown a) with respect to CuO$_6$ octahedra,
 b) within $Bmab$ unit cell of La$_2$CuO$_4$ 
 (green - Cu atoms, grey - La atoms, blue - O atoms, 
 red - muon positions).
}
\end{center}
\vspace*{-8mm}
\end{figure}

We determine
the muon stopping sites 
using the 
dipole-field approximation and assuming the 
periodic AFM structure of La$_2$CuO$_4$ 
with the known moments on Cu atoms.
High-field $\mu $SR spectra were recorded 
for three different mutually orthogonal 
orientations of the sample, 
providing 
absolute values of the local magnetic field 
projections 
at the muon sites.  
There are two sites: 
the small-amplitude, small-field site gives the inner two peaks in TF 
(one for each direction of the AFM dipolar field); 
the large-amplitude, large-field site gives the outer peaks 
(each split by some additional contribution).  
Fig.~\ref{fig:La2CuO4-muons} shows 
chemically possible muon positions
consistent with the $\mu $SR data.  
The picture is quite simple: the muon is bound 
to an apical oxygen atom, but it can be 
located on the side closer to a Cu atom 
(large amplitude signal with higher ZF frequency) 
or the other side, closer to a La atom 
(small amplitude signal with lower ZF frequency).

The muon is located somewhat closer to 
the oxygen atom 
than expected from the typical 
O-H bond 
distance of $\sim 1$~\AA. 
This can be explained by a 
small 
perturbation of the loosely bound 
apical oxygen 
position due to its bonding with the muon. 
Other corrections ({\it e.g.} 
uncertainty in the 
value of the magnetic moment on Cu 
and 
approximations of the computational 
procedure) are estimated to be small and 
therefore insignificant to the general conclusions.

The magnetic fields are almost indifferent 
to the tilting of CuO$_6$ octahedra because 
the canted component of the moment on Cu is very small. 
Nevertheless, the tilting produces two types 
of structurally inequivalent positions for muons 
differing in the distance between the 
muon and the apical oxygen atom: 
shortened (SD) and elongated (ED) 
(Fig.~\ref{fig:La2CuO4-muons} shows ED muon positions). 
The ED position seems preferable, as it 
produces less perturbations on oxygen. 
In any case, the choice between ED and 
SD does not affect our conclusions.

\begin{figure}[tb]
\begin{center}
\vspace*{-2mm}
\includegraphics[width=0.8\columnwidth,angle=0]{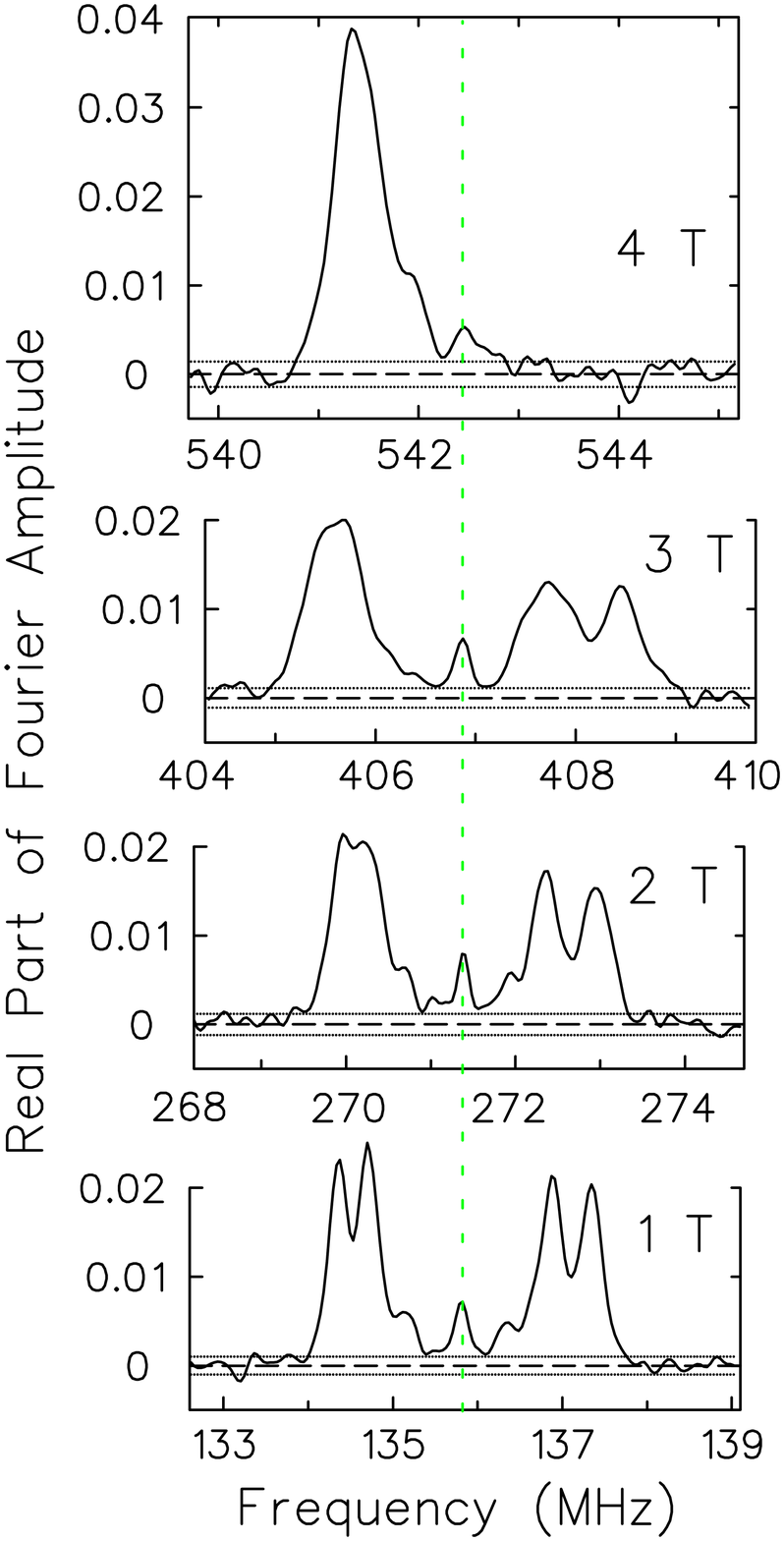}
\vspace*{-4mm}
\caption{\label{fig:La2CuO4-cparl-FFT-250K}
 Fourier transforms of the muon spin precession signals
 in AFM stoichiometric La$_2$CuO$_4$ at $T=250$~K
 in different external magnetic fields 
 directed along the $\hat{c}$-axis of the crystal. 
 The central line corresponds to the background signal.  
}
\end{center}
\vspace*{-7mm}
\end{figure}

The observed splittings cannot be 
explained by different muon sites: 
In high magnetic field applied 
along the $\hat{c}$-axis, a spin-flop 
transition to weak ferromagnetism 
takes place \cite{Kastner1998,Thio1988}. 
The direction of small ferromagnetic 
moments in every CuO$_2$ plane becomes 
the same and is accompanied by 
reversal of in-plane components of 
magnetic moments for every second CuO$_2$ plane. 
At $T=250$~K, SQUID measurements of our samples show 
that a 
spin-flop transition occurs in a magnetic field $H \sim 4$~T \cite{Storchak2014}. 
The jump of magnetization $\sim 0.002$~$\mu_{\rm\scriptsize B}$ 
is consistent with the literature data 
\cite{Kastner1998,Thio1988}. 
Such a transition makes the magnetic unit cell equivalent 
to the crystallographic unit cell and the number of magnetic 
sublattices decreases from four to two.  
This spin-flop transition is clearly seen in 
$\mu $SR measurements at 250 K 
(see Fig.~\ref{fig:La2CuO4-cparl-FFT-250K}) 
where half of the lines in the spectra 
vanish in a magnetic field of about 4~T. 
A similar feature does not show up
in $\mu $SR spectra
at 50~K 
as the critical field of spin-flop 
transition depends strongly on the 
temperature and the doping level 
of La$_2$CuO$_4$: 
A sample with $T_N\sim 234$~K has $H_c\sim 3$~T at 
200~K \cite{Thio1988} while a sample 
with $T_N \sim 316$~K has $H_c \sim 12$~T at 
1.8~K \cite{Reehuis2006}. 
The changes in the spectra 
verify that 
{\sl muons occupy only half of the possible 
crystallographic positions\/} (either ED or SD).

Since the splitting of the large amplitude 
lines cannot be explained by the inequivalence 
of crystallographic positions brought on by 
tilting of the  CuO$_6$ octahedra, one can try 
to ascribe it to structural deformations 
leading to two types of tilted octahedra. 
Indeed, there are suggestions that a lower 
symmetry ($Bm11$ space group) is possible for 
La$_2$CuO$_4$, but our calculations show that 
the proposed structural deformations are more 
than 3 times smaller than the difference in 
muon positions required to explain 
such a large splitting of the lines. 
Moreover, equal amplitudes of the split 
signals 
(see Fig.~\ref{fig:La2CuO4-100K-1T-spec}) 
would be a surprising coincidence for different 
muon stopping sites. 

Therefore we conclude 
that the splittings are due to an additional 
source of the magnetic field (besides AFM). 
This field can not originate from the apical oxygens 
because it should cause comparable splittings 
of both low- and high-frequency signals. 
Instead, the source of this additional magnetic field 
is likely confined 
within CuO$_2$ planes. 
Then the apparent absence of splittings 
in the low-frequency signals can be 
easily 
explained 
by the increased distance from the muon 
to CuO$_2$ planes, leading to much smaller 
splittings 
that are not resolved in 
the experiment.

There are two major types of models 
explaining intra-unit-cell magnetic order in cuprates. 
One is based on spin polarization of oxygen atoms 
while the other relies upon circulating orbital currents. 
Polarized neutron experiments have not been able to differentiate 
the existing models beyond establishing TRSB 
\cite{Fauque2006,Bourges2008}. 
In contrast, high-field $\mu $SR experiments 
(in combination with simple symmetry considerations) 
may provide such differentiation 
as the ratio of splittings for different directions 
of the magnetic field differs for different models.

\begin{figure}[tb]
\begin{center}
\vspace*{3mm}
\includegraphics[width=\columnwidth,angle=0]{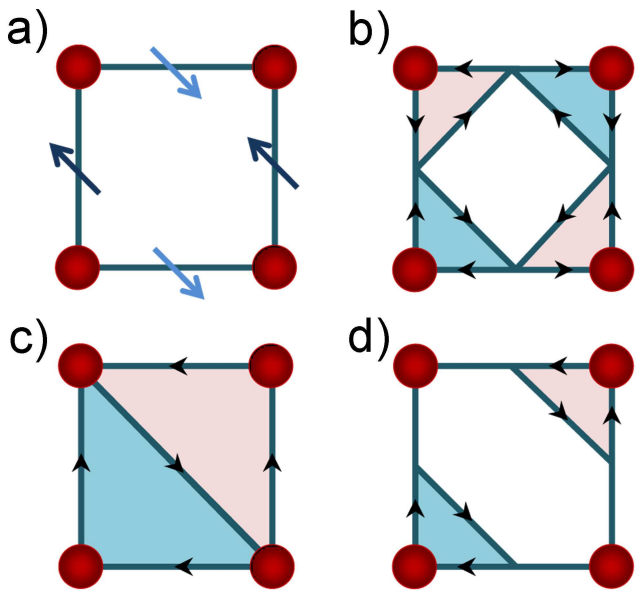}
\vspace*{-6mm}
\caption{\label{fig:La2CuO4-order}
 Proposed sources of magnetic order in La$_2$CuO$_4$: 
 a) AFM order of spins on oxygen atoms;
 b) $CC$-$\Theta _I$ model with 4 orbital current loops per CuO$_2$ plaquette;
 c) staggered orbital current phase with Cu-Cu-Cu current loops;
 d) $CC$-$\Theta _{II}$ model with two opposite current loops 
    O-Cu-O per 
    plaquette.
}
\end{center}
\vspace*{-9mm}
\end{figure}

The first class of models is based on the 
AFM ordering of spins on 
in-plane oxygen atoms 
(see Fig.~\ref{fig:La2CuO4-order}a). 
When the spins are directed along the $\hat{a}$-axis 
\cite{Fauque2006} the model has the correct 
symmetry structure to describe 
the splittings in the $\mu $SR spectra, but 
the ratio of splittings along the $\hat{b}$- and 
$\hat{c}$-axes is about 1.6, while our experiments 
determine it to be close to 1.0. 
More importantly, this model predicts large 
($\sim 40$~G) splitting of the 
high-frequency signal in the absence of 
magnetic fields, which is not observed in the 
ZF-$\mu $SR spectra. 
An alternative model with antiferromagnetic ordering of 
spins directed along the $\hat{c}$-axis 
\cite{Moskvin2012} has an incorrect symmetry 
structure 
since 
it does not provide 
splitting of signals for fields along the $\hat{c}$-axis.

Alternatively, magnetic order can be based on current 
loops 
within the unit cell. The original 
model $CC$-$\Theta _I$ (Fig.~\ref{fig:La2CuO4-order}b) is 
formed by 4 orbital current loops O-Cu-O 
for each CuO$_2$ plaquette \cite{Varma1997}. 
This model is not consistent with 
either 
polarized neutron 
experiments 
or 
our $\mu $SR data, 
as the symmetry of the model prevents 
the splitting of signals for fields directed along the $\hat{c}$-axis. 
An alternative one-band model with 
staggered orbital current phase 
formed by Cu-Cu-Cu currents 
\cite{Stanescu2004} 
(Fig.~\ref{fig:La2CuO4-order}c) has 
correct symmetry properties but the 
calculated ratio of splittings 
along the $\hat{b}$- and $\hat{c}$-axes is too large 
and the absence of splitting in the ZF spectra 
can not be explained within this model.

The most widely used CC model of 
intra-unit-cell magnetic order is $CC$-$\Theta _{II}$ 
(Fig.~\ref{fig:La2CuO4-order}d) 
with two opposite current loops O-Cu-O 
per CuO$_2$ plaquette \cite{Varma2006}. 
The basic symmetry properties are correct 
and the calculated ratio of splittings 
along the $\hat{b}$- and $\hat{c}$-axes ($\sim 1.1$) 
is close to the experimental value, 
provided that the moments follow the 
natural tilting of CuO$_6$ octahedra 
and the muon occupies ED positions. 
However, this model still provides 
an incorrect description of 
ZF-$\mu $SR spectra.

Polarized neutron experiments 
reveal that for different families of 
cuprates the moments responsible for 
the intra-unit-cell magnetic order should 
be significantly tilted, with a tilting 
angle 45$\pm $20$^\circ $ \cite{Fauque2006,Bourges2008}. 
This constitutes a significant departure from the original 
CC model predicting the orbital moments to 
be orthogonal to CuO$_2$ planes. 
However, attempts to reconcile the theory and the experiment 
have been made 
based on accounts of spin-orbit interactions \cite{Aji2007} 
and quantum interference of current loop states \cite{He2011}. 
Our calculations show that the correct ratio of splittings 
is achieved for a tilting of orbital moments by 
$\sim 53^\circ$ within the $CC$-$\Theta _{II}$ model. 
Moreover, this modified model is the only one consistent with
zero magnetic field data:
the changes of the magnetic field have opposite signs 
for the in-plane component and that along the $\hat{c}$-axis; 
thus 
the splitting of the ZF signal 
becomes small and 
hence 
not resolved in 
ZF-$\mu $SR experiments. 
The orbital moment extracted from our data amounts to 
about 0.04~$\mu_{\rm\scriptsize B}$, which is comparable with 
the theoretical value on the order of 
0.1~$\mu_{\rm\scriptsize B}$ \cite{Varma1997}.

One puzzling characteristic of the spectra must 
be 
mentioned.
In high magnetic field parallel to the $\hat{c}$-axis,  
the splitting of the component antiparallel to the 
field decreases while the splitting of the component 
parallel to the field increases; 
the spin-flop transition 
decreases the splitting. 
This 
effect is not detected 
with 
magnetic fields applied along the other axes.
This peculiarity can be 
phenomenologically 
described if one assumes 
modulation of the tilting angle for the orbital magnetic moments 
due to their coupling to the canting of moments on Cu. 
However, 
we see no apparent reason for such coupling,
which calls for additional studies.  

In summary, high magnetic field 
$\mu $SR studies in 
orthorhombic La$_2$CuO$_4$ reveal an 
additional source of magnetic field 
within the unit cell,
consistent with the Varma model of 
circulating currents in cuprates. 

This work was supported by 
Kurchatov Institute, 
NSERC of Canada 
and the U.S. DOE, Basic Energy Sciences (Grant DE-SC0001769).
\vspace*{-2mm}   

\end{document}